\documentclass[journal]{IEEEtran}
\usepackage{cite}
\usepackage{amsmath,amssymb,amsfonts}
\usepackage{algorithmic}
\usepackage{graphicx}
\usepackage{textcomp}
\def\BibTeX{{\rm B\kern-.05em{\sc i\kern-.025em b}\kern-.08em
    T\kern-.1667em\lower.7ex\hbox{E}\kern-.125emX}}

\usepackage{graphicx}
\usepackage[export]{adjustbox}
\usepackage{amssymb}
\usepackage[utf8]{inputenc}
\usepackage[ nomarginpar, top=0.8in, bottom=1.25in, left=1in, right=1in]{geometry}
\usepackage{siunitx}
\usepackage{lscape}
\usepackage{makecell}
\usepackage{booktabs}
\usepackage{verbatim}
\usepackage{url}

\title{Is Your Smartband Smart Enough to Know Who You Are: Continuous Physiological Authentication in The Wild}

\begin{document}
\author{Deniz~Ekiz, 
        Yekta Said~Can,
        ~Yağmur Ceren~Dardağan
         
         and  ~Cem~Ersoy
        
\IEEEcompsocitemizethanks{\IEEEcompsocthanksitem Deniz Ekiz, Yekta Said Can, Yağmur Ceren Dardağan and Cem Ersoy were with the Department
of Computer Engineering, Boğaziçi University, Turkey.\protect\\
E-mail: deniz.ekiz@boun.edu.tr 
}
\thanks{Manuscript received XXX, XXX}}

%



\IEEEtitleabstractindextext{
\begin{abstract}
The use of cloud services that process privacy-sensitive information such as digital banking, pervasive healthcare, smart home applications requires an implicit continuous authentication solution which will make these systems less vulnerable to the spoofing attacks. Physiological signals can be used for continuous authentication due to their personal uniqueness. Ubiquitous wrist-worn wearable devices are equipped with photoplethysmogram sensors which enable to extract heart rate variability (HRV) features. In this study, we show that these devices can be used for continuous physiological authentication, for enhancing the security of the cloud, edge services, and IoT devices. A system that is suitable for the smartband framework comes with new challenges such as relatively low signal quality and artifacts due to placement which were not encountered in full lead electrocardiogram systems. After the artifact removal, cleaned physiological signals are fed to the machine learning algorithms. In order to train our machine learning models, we collected physiological data using off-the-shelf smartbands and smartwatches in a real-life event.  Performance evaluation of selected machine learning algorithms shows that HRV is a strong candidate for continuous unobtrusive implicit physiological authentication.
\end{abstract}

}

\maketitle

\IEEEdisplaynontitleabstractindextext

%
\IEEEpeerreviewmaketitle

\section{Introduction}
\label{S:1}

%
%
%
%

\IEEEPARstart{I}{mplicit} continuous authentication is required for cloud oriented services which grant access to the privacy sensitive information domains such as mobile banking, pervasive healthcare \cite{Baktir18,Ntantogian2015}.  Smartphones, computers, smartwatches and Internet-of-Things (IoT) devices become more dependent on these services. It is expected that the number of IoT devices will be more than 75 Billion in 2025 \cite{iotstatis}. However, these services are vulnerable to attacks once users authenticate. For example, a smartphone can be forgotten logged-in, the privacy-sensitive services and information can be stolen by the attackers.  One simple mechanism can be asking a password to the user frequently.  However, this is annoying for the service users. Continuous authentication should be implicit. Face-based systems can be tricked by using presence attacks, such as printing the face of the victim on a paper. Storing the face pictures of the users also create a privacy concern \cite{surveyaccess}. Furthermore, fingerprint which is another prominent traditional biometrics modality along with the face-based systems can be easily manipulated \cite{ross2008introduction} and fails on liveness detection tests. On the other hand, biosignals are difficult to temper with and they inherently have liveness detection feature \cite{akhter2015heart}. Heart activity is unique to individuals and biosignal authentication research has started investigating this signal \cite{akhter2015heart}. One of the most important properties of this signal is the Heart Rate Variability (HRV). Although research in the past mostly focused on the connection between HRV and different types of health disorders \cite{akhter2012microcontroller}, the validity of using HRV for biometric recognition is supported by the fact that the physiological and geometrical differences of the heart in different individuals display certain uniqueness in their HRV features \cite{Ramli2016}. \\
\indent High-end wearable systems are expensive and provide low comfort for the users, which limit their wide range application. Recently, smart bands and smartwatches became widely adopted by consumers. These devices are equipped with a rich set of sensors such as accelerometer, heart rate monitor and skin conductance. These advances create an opportunity to build a continuous implicit authentication system. However, these devices are prune to activity related errors \cite{Can2019} unlike full lead ECG systems. Modality specific artifact detection and removal mechanisms should be developed for accurate measurements. A solution suitable for IoT connected devices should be context-independent because every service may require different types of behavior. Therefore, systems that only work in certain scenarios such as while typing or walking are very limited in terms of the application area.  The physiological parameters like hidden heart-related biometrics are more suitable for this purpose due to their uniqueness and activity independence \cite{phua2008heart}, \cite{pirbhulal2015efficient}. We propose an unobtrusive, low cost, activity-independent continuous authentication system with smartbands. We implemented our solution on Empatica E4 Smartband \cite{empatica}, Samsung Gear S  and Gear S2 \cite{Samsung} which are equipped with a photoplethysmography (PPG) sensor.

Let's think about a scenario, where John has to make money transactions through a mobile banking smartphone application. He logins to the system via two-factor authentication. After he successfully makes the transaction, he forgets to logout. Then, an attacker withdraws all the cash from John's account by using his session.  If John has a continuous biometric authentication system, he would not have this problem. Such a system can be also used in services like city bikes and electric scooter rental services which have become popular in recent years.  The literature on continuous authentication with physiological signals gathered from smartbands is limited. The effectiveness of HRV features derived from PPG sensors of smartwatches and smartbands is still unknown for continuous authentication \cite{surveyaccess}. Our proposed system enhances the state of the art as follows:
 \begin{itemize}
      \item One of the previous studies employed mean heart rate per minute which requires longer recordings and achieves approximately five minutes \cite{Vhaduri2019}. The proposed system uses more sophisticated HRV features derived from inter-beat intervals which are extracted from the raw PPG sensor data. 
     \item Most of the previous works with smartbands only focused on continuous authentication while using a laptop, a smartphone, an ATM and used lifestyle related metrics. Our solution is not activity dependent. For example, a user may use this system during any activity such as live streaming,  on social media, cycling, presentation or working in an office.
     \item Our solution is comfortable, unobtrusive, seamless and works without interrupting the ordinary pattern of any activity.
      \item Real-life data contains artifacts. We evaluate the effect of data quality on system performance.
 \end{itemize}

In Section \ref{background} we provide background information on the smartwatch framework and heart rate variability. In Section \ref{relatedwork}, the related work on continuous authentication and the comparison with our work in terms of  its novelty are presented. In Section \ref{system}, we describe the proposed system for continuous authentication with smartwatches using heart rate variability. In Section \ref{experiment}, we explain the conducted experiment for the proposed system. In Section \ref{results}, we provide the results of our system. In Section \ref{conclusion}, we present the conclusion and future works of our study.

\section{Background}
\label{background}
\subsection{Smartband Framework}

\begin{figure}[hb!]
    \centering
    \includegraphics[width= \columnwidth]{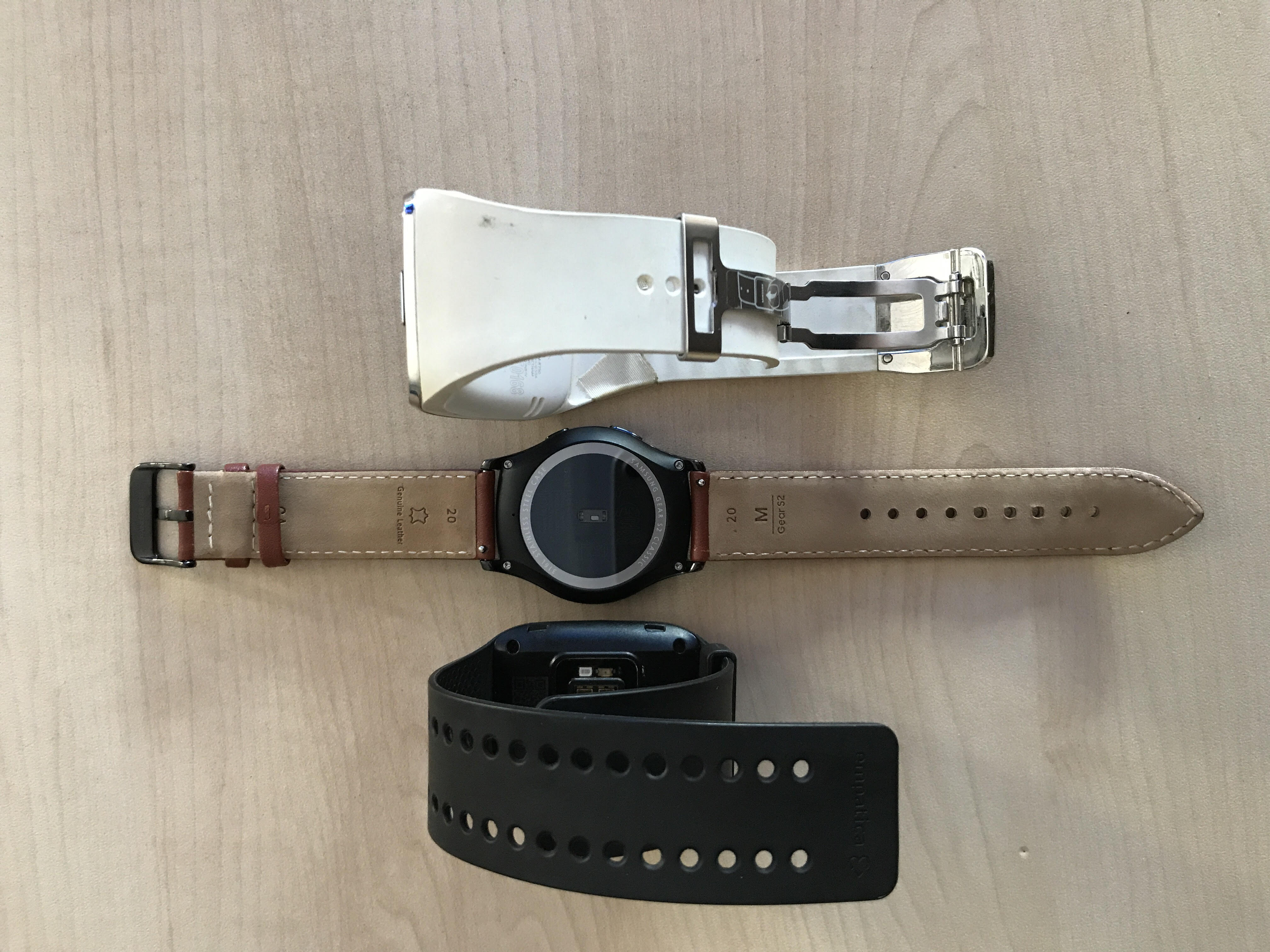}
    \caption{Empatica E4, Samsung Gear S2 and Samsung Gear S are shown in the order.}
    \label{fig:straps}
\end{figure}

Smartbands (some times called wristbands or smartwatches) are comfortable devices that can be attached to the wrist or arm. The devices used in this study are shown in Figure \ref{fig:straps}.
Recently, the battery life of smartbands is increased to days. For example, the battery life of Empatica E4 is two days when all sensors are used \cite{empatica}. The extended battery life of smartbands enables monitoring physiological signals gathered from individuals for long periods. Most of the wristbands are equipped with PPG which is an optically obtained signal that can be used to detect blood volume changes in the microvascular bed of tissue \cite{Sun2016}. From the PPG signal, the time between the beats (RR intervals) can be computed. Most of the modern smartbands provide RR intervals thanks to their APIs. For example, Samsung smartwatches use the Tizen framework which has Human Activity Monitor API to gather the RR intervals. The sampling frequency of the PPG sensor can vary from 20Hz to 100Hz (64Hz for E4, 100Hz for Samsung Gear Series) in many smartbands. Cubic spline interpolation is used to detect the beats more accurately and most of the devices correct the heartbeats by using an accelerometer sensor for detecting the movements of the subject. This functionality is available in Tizen and Empatica API \cite{empatica}. These devices are also equipped with Bluetooth and recently NFC chips which enable them to connect to smartphones, edge and cloud services.  These short-range network interfaces can be used to check if the user is in the close proximity of the computer they are using. 

\subsection{Heart Rate Variability}

The variability of RR intervals is called HRV \cite{hrvguideline}.  It is a very important feature for recognition of certain psychological, physiological and personal properties of an individual \cite{hrvguideline}. In the literature, Kubios \cite{kubios} is a popular HRV feature extraction tool to compute the HRV.  Non-linear, time domain and frequency domain features can define the variability of the heart \cite{kubios}. The calculation of frequency domain and time domain features of HRV is computationally effective, thanks to Fast Fourier Transform (FFT) $O(n\log{}n)$ for frequency domain and $O(n)$ for time-domain features.

\section{Related Work}
\label{relatedwork}

\begin{table*}[htb!]
\centering
\label{tab:relatedwork}
\footnotesize
\caption{Previous works on continuous authentication.}
\begin{adjustbox}{width=\textwidth,center}

   \begin{tabular}{ | c c c c c c c c | }
 
  \hline

\textbf{\makecell{Article}} &
\textbf{\makecell{Year}} & 
\textbf{\makecell{Device(s)}} &
\textbf{\makecell{Device Position(s)}} & 
\textbf{\makecell{Features}} &
\textbf{\makecell{Unobtrusive}} &
\textbf{\makecell{Environment}} &
\textbf{\makecell{Activity \\ Independent}}  \\
\hline

{ \makecell{S. A. Elkader \\et al. \cite{Elkader2018}}}  &{ \makecell{2018}}  &{ \makecell{A combination of\\ acceleration,\\ gyroscope and \\magnetometer \\ sensors}}  &{ \makecell{Dominant wrist,\\ dominant upper\\arm, non-\\dominant\\wrist,thigh, \\chest, ankle}}  &{ \makecell{Accelerometers,\\ gyroscope and \\magnetometers \\ features extracted \\from six \\sensor-based \\body locations}}  &{ \makecell{No}} & { \makecell{ Laboratory \\ Kitchen area}}  &{ \makecell{No}}  \\
  
{ \makecell{Y. Zeng \\et al. \cite{Zeng2017}}} & { \makecell{2017}} & { \makecell{Shimmer 6DoF \\IMU \\ Samsung Galaxy \\S4 i9500 \\ Samsung Galaxy\\ Nexus i9500}} &{ \makecell{Wrist,ankle,\\ hip/torso, \\ thigh/front \\pocket,\\ upper arm}} & { \makecell{Statistical features \\ extracted from \\ accelerometer sensor} }& { \makecell{Yes}} & { \makecell{Laboratory}} &{ \makecell{No}}\\ 

{ \makecell{S. Vhaduri\\ et al. \cite{Vhaduri2017}}} & { \makecell{2017}} & { \makecell{Fitbit Charge \\HR device}} &{ \makecell{Wrist}} & { \makecell {Statistical features \\ extracted from  \\step counts, heart rate, \\ calorie burn, \\ metabolic equivalent of\\ task information }}& { \makecell{Yes}} & { \makecell{Daily Life}} & { \makecell{No}} \\ 

{ \makecell{S. Vhaduri\\ et al. \cite{Vhaduri2019} } }& { \makecell{2019}} & { \makecell{Fitbit Charge  \\HR device}} &{ \makecell{Wrist}} & { \makecell{Statistical features \\ extracted from  \\step counts, heart rate, \\ calorie burn, \\ metabolic equivalent of\\ task information }}&{ \makecell{Yes}}& { \makecell{Daily Life}} &{ \makecell{No}} \\ 

{ \makecell{Matsuyama \\ et al. \cite{Matsuyama2015}}} & { \makecell{2015}} &{ \makecell{Near-infrared \\ Spectroscopy \\ (NIRS)}} &{ \makecell{Fore Head}} &{ \makecell{Low-frequency \\ brainwaves}} &{ \makecell{No}} &{ \makecell{Laboratory}} &{ \makecell{No}}  \\

{ \makecell{Sarkar \\et al. \cite{Sarkar2016_2}}} & { \makecell{2016}} &{ \makecell{Front facing \\ camera of \\ mobile device}} &{ \makecell{In front of \\ face }} &{ \makecell{Deep Face Features}} &{ \makecell{Yes}} &{ \makecell{Daily Life}} &{ \makecell{No}}  \\

{ \makecell{Ramli \\ et al. \cite{Ramli2016}}} & { \makecell{2016}} &{ \makecell{Hearbeat\\ detection \\ kit as a \\ wearable \\ bracelet}} &{ \makecell{Wrist}} &{ \makecell{ECG Wavelet\\ Features}} &{ \makecell{No}} &{ \makecell{Laboratory}} &{ \makecell{No}}  \\

{ \makecell{Ntantogian \\et al. \cite{Ntantogian2015}}} & { \makecell{2015}} &{ \makecell{Camera}} &{ \makecell{In front of \\ a person }} &{ \makecell{ Gait  and gesture features }} &{ \makecell{Yes}} &{ \makecell{Laboratory}} &{ \makecell{No}}  \\

{ \makecell{Musale \\et al. \cite{Musale2019}}} & { \makecell{2019}} &{ \makecell{Motorola \\ 360 Sport \\2nd Gen \\
(smartwatch)\\ Motorola \\G4 plus \\(smartphone)}} &{ \makecell{Wrist}} &{ \makecell{ Statistical and \\ human-action-related \\ features from \\ accelerometer and \\gyroscope sensor }} &{ \makecell{Yes}} &{ \makecell{Daily Life}} &{ \makecell{No}}  \\

{ \makecell{Peng \\et al. \cite{Peng2017}}} & { \makecell{2017}} &{ \makecell{Google Glass}} &{ \makecell{Head}} &{ \makecell{ Touch gestures (single-tap,\\ swipe forward, swipe \\backward, swipe down,\\ two-finger, swipe \\forward, and two-finger \\swipe backward) and \\voice commands }} &{ \makecell{No}} &{ \makecell{Laboratory}} &{ \makecell{No}}  \\

{ \makecell{Our Work}} & { \makecell{2019}} &{ \makecell{ Samsung Gear S \\ Samsung Gear S2 \\ Empatica E4}} &{ \makecell{Wrist }} &{ \makecell{ Heart Rate Variability  \\ features derived \\ from PPG sensor }} &{ \makecell{Yes}} &{ \makecell{Real Life}} &{ \makecell{Yes}}  \\

  \hline
  \end{tabular}
  \label{tab:relatedwork}

  \end{adjustbox}
\end{table*}

The behavioral and physiological biometrics from the wearable devices via sensors have become popular in individual recognition and authentication models. Some models focus on biometrics such as face \cite{Ghayoumi2015}, voice \cite{Brunelli1995}, fingerprints \cite{Camlikaya2008}, \cite{Hammad2019}, electroencephalography (EEG)   \cite{Bugdol2014}, \cite{Gui2014}, \cite{Rangoussi2002}, ECG \cite{ZHANG2017}, \cite{Choi2016}, \cite{Hammad2019} and phonocardiography (PCG), \cite{Sarkar2016}. The continuous authentication field is a fast-growing field, however, the literature on a system that is not dependent on a certain type of task is limited. Most of the previous work using physiological signals have been done on laboratory-grade equipment. Some of these sensors are not available in unobtrusive devices such as smartphones, smartwatches, and smart bands. On the other hand, fingerprints and face-based authentication systems can be easily deceived. Authentication with voice has privacy issues, which requires continuous voice recording of the environment.

As a method for recognizing individuals, Elkader et al. \cite{Elkader2018} presented a sensor-based motion biometric model that is suitable for 20 sedentary and non-sedentary activities (Vacuuming, Sweeping, Walking Downstairs, Walking Upstairs, Dusting, Iron Cloth, Folding Cloth, Washing Hands, Brushing Teeth, Washing Dishes, Washing Vegetables, Dicing, Peeling Vegetables, Grating, Stirring, Watching TV, Using PC, Talking on Phone Texting on Phone, Writing with Pen). They used different combinations of 3 sensors (acceleration, gyroscope and magnetometer sensors) on 6 different body positions (dominant wrist, dominant upper arm, non-dominant wrist, chest, thigh, and ankle positions).  They concluded that features extracted from the combination of six sensors reach the best classification accuracy in overall (98.3\%). These activities are gathered in a laboratory environment with manual segmentation of the signals.

Another approach for implicit identification and authentication based on activity information, WearAI, Zeng et al.\cite{Zeng2017}, proposed a biometric model that utilizes accelerometer and gyroscope sensors from five body locations such as left wrist (Shimmer 6DoF IMU), right ankle (Shimmer 6DoF IMU), center right hip/torso (Samsung Galaxy S4 i9500), left thigh/front pocket (Samsung Galaxy Nexus i9250), right upper arm (Samsung Galaxy Nexus i9250)). They achieved 97\% accuracy with less than 1\% false-positive rate. However, in both methods, placing many sensors on the body can be disturbing for the user in daily life usage.

Acar et al. \cite{Acar18} used smartwatches with keystroke dynamics for continuous authentication.
Musale et al. \cite{Musale2019} proposed a continuous authentication system based on Motorola 360 Sport by using accelerometer and gyroscope features.
Vhaduri et al. \cite{Vhaduri2017} proposed continuous user authentication scheme that uses 44 features extracted from various biometrics (calorie burn, metabolic equivalent of task (MET), heart rate and step count) using Fitbit Charge HR device and they achieved average accuracy of 87.37\% with Quadratic SVM classifier in one-to-many approach and average accuracy of 93\% with Quadratic SVM classifier in one-to-one approach. In their revised scheme \cite{Vhaduri2019}, they adopted more features (65) with different feature selection approaches and 93\% (sedentary) and 90\% (non-sedentary) with equal error rates of 5\% is obtained. However, the Fitbit framework only provides only one sample each minute and access to the raw data is not possible.  A system for continuous authentication with physiological signals should be low-cost and unobtrusive, and should not be dependent on certain activity for the sake of universality. We compared the proposed system with the related work in Table \ref{tab:relatedwork} in terms of device and device position, features, unobtrusiveness, environment, and dependency to the activity type. Our system outperforms other studies when feature engineering complexity, activity independence and unobtrusiveness are taken into consideration.

\begin{figure}[htb!]
    \centering

    \includegraphics[width=\columnwidth]{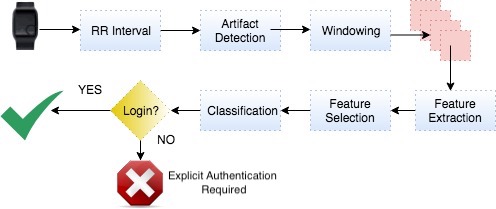}
    \caption{The system diagram of our proposed solution.}
    \label{fig:authentication}
\end{figure} 

\begin{figure}[hb!]
    \centering

    \includegraphics[width=0.5\textwidth]{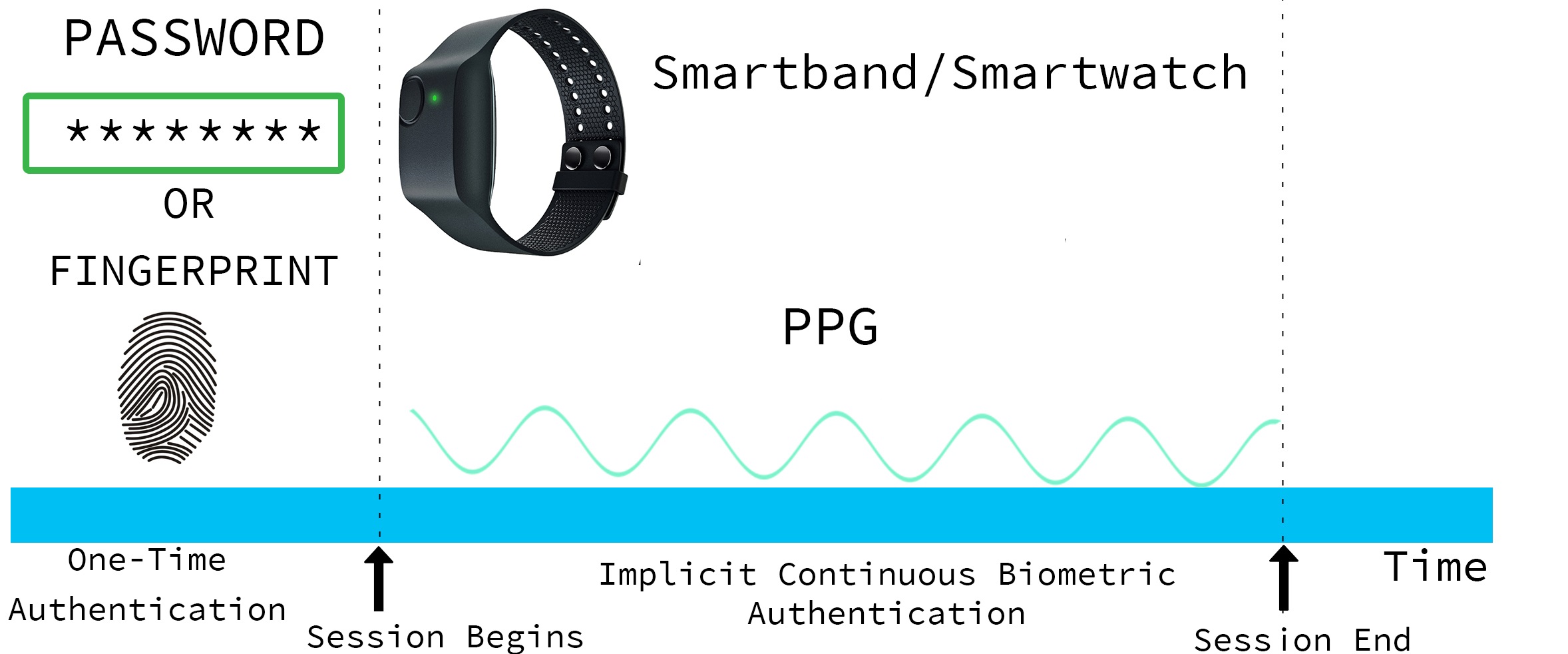}
    \caption{Our proposed solution continuously authenticate the user by processing the RR intervals coming from the smartband.}
    \label{fig:system}
\end{figure}

\section{Proposed System}

\label{system}

In this section, we explain our continuous authentication scheme. In Figure \ref{fig:authentication}, we show the data collection application, preprocessing for artifact detection, feature extraction, feature selection and classification units of our system. The overall multi-factor authentication diagram where a user initiates his/her session with a password, or fingerprint, is shown in Figure \ref{fig:system}.

\subsection{Data Collection Application}

\begin{figure}[hb!]
    \centering
    \includegraphics[width=0.5\textwidth]{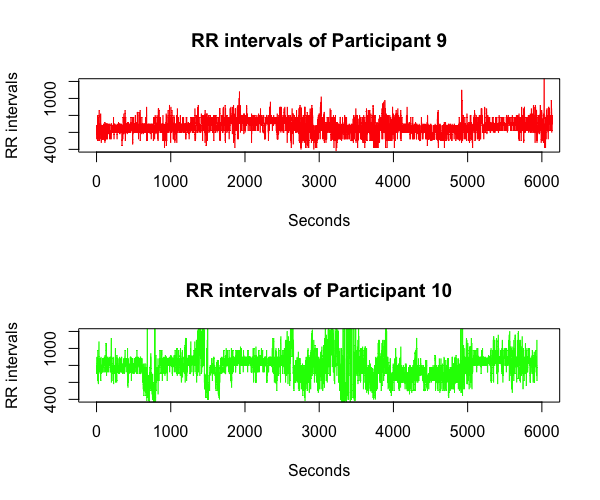}
    \caption{The RR intervals gathered from two participants using Samsung Gear S. The difference can be seen by looking at the raw signals.}
    \label{fig:comparision}
\end{figure}

  We developed a data collection application in Tizen 3.0 framework \cite{tizen} for Samsung Gear S and S2. The application collects inter-beat intervals and 3D accelerometer data and stores them as downloadable comma-separated values (CSV files).  Empatica E4 has a cloud based data collection application. The physiological signals can be downloaded as a CSV file.   The gathered RR intervals from two different participants are shown in Figure \ref{fig:comparision}.

\subsection{Preprocessing and Artifact Removal}

We implemented our preprocessing module in MATLAB \cite{MATLAB:2010}. First, we loaded the CSV file provided from the smartbands. The signal is segmented into non-overlapping time windows of 120 seconds. According to the HRV guidelines, 2 minutes is the minimum window length for calculation of short-term HRV features \cite{hrvguideline}. Since response time is important for a continuous authentication system, we selected the minimum possible duration. Therefore, the minimum required duration of physiological data for authentication is 2 minutes. The artifacts in the RR intervals are detected by checking the difference between the consecutive points. We labeled the points exceeding more than 20\% of the local average as artifacts, and the other points as the validated RR intervals, this threshold is selected from the previous works \cite{Cinaz2013}. The points labeled as artifacts are deleted.  After the removal, we implemented two different techniques. The first one is to interpolate the missing data points using a cubic spline interpolation algorithm which is commonly used \cite{kubios}. The second technique is to apply the minimum consecutive time and sample constraints on the remaining data to be regarded as meaningful. For example, if the minimum sample constraint is set to 5, we do not count three consecutive samples followed by a missing data point because the sequence is too short to be evaluated. In this study, we applied the former technique because it achieved better results \cite{Can2019}.

\subsection{Feature Extraction}

We extracted time and frequency domain heart rate variability features from the segmented time windows. We used Marcus Volmer's toolbox \cite{MarcusVollmer} which is implemented in MATLAB. We selected the features which are commonly used in the previous works related to heart rate variability \cite{Can2019}, \cite{Cinaz2013} and \cite{gjoreski2017monitoring}. In order to compute the frequency domain features, the RR intervals are interpolated using 4Hz cubic spline interpolation, because RR intervals are unevenly sampled. We applied FFT to the interpolated windows. The computed features are shown in Table \ref{table:hrv_fatures}. The total number of extracted features is 11 for each window.

\begin{table}[h!]
\centering

\caption{Heart rate variability features and their definitions.}
\label{table:hrv_fatures}
\begin{tabular}{|p{2.5cm} |p{3.5cm}|}
\hline
\textbf{HRV Feature}     & \textbf{Description}                                                                                                                                \\ \hline
Mean RR              & Mean value of the RR intervals                                                                                                                                 \\ \hline
STD RR               & Standard deviation of the RR intervals                                                                                                                              \\ \hline
RMSSD                & Root mean square of successive difference of the RR intervals                                                                                                               \\ \hline
pNN50                & Percentage of the number of successive RR intervals varying more than 50ms from the previous interval                           \\ \hline
HRV triangular index & Total number of RR intervals divided by the height of the histogram of all RR intervals  measured on a scale with bins of 1/128 s\\  \hline
TINN                 & Triangular interpolation of RR interval histogram                                                                                                                                      \\ \hline
SDSD                 & Related standard deviation of successive RR interval differences                                                     \\ \hline
LF                   & Power in low-frequency band (0.04-0.15 Hz)                                                                                                                                  \\ \hline
HF                   & Power in high-frequency band (0.15-0.4 Hz)                                                                                                                                  \\  \hline
LF/HF                & Ratio of LF-to-HF                                                                                                                                                           \\ \hline
VLF                  & Power in very low-frequency band (0.00-0.04 Hz)                                                                                                                                                                         \\ \hline
\end{tabular}
\end{table}

\subsection{Dimensionality Reduction}

It is known that dimensionality reduction leads to better performance for the machine learning systems since it removes the unrelated features with the desired prediction \cite{barberBRML2012}.
We used Principal Component Analysis (PCA) based dimensionality reduction which is available in the Weka toolkit \cite{eibe2016weka}. PCA is a very powerful tool when applied with machine learning (ML) classifiers. It converts the set of vectors to uncorrelated variables. We explored the effect of different selection of covered variance and PCA variables. It is known that the covered variance affects the classification performance, therefore different values are evaluated (0.8, 0.85, 0.9 and 0.95) we reported the best results when the variance is selected as 0.9.

\subsection{Handling Class Imbalance}
Since some of the windows are deleted due to improper placement of the devices or heavy movements. There is a class imbalance between participants. We applied the majority class subsampling to equalize the number of windows for each participant. This method is the most commonly used one in the literature \cite{undersample_rg}.

\subsection{Machine Learning}

\begin{figure}
    \centering
    \includegraphics[width=\columnwidth]{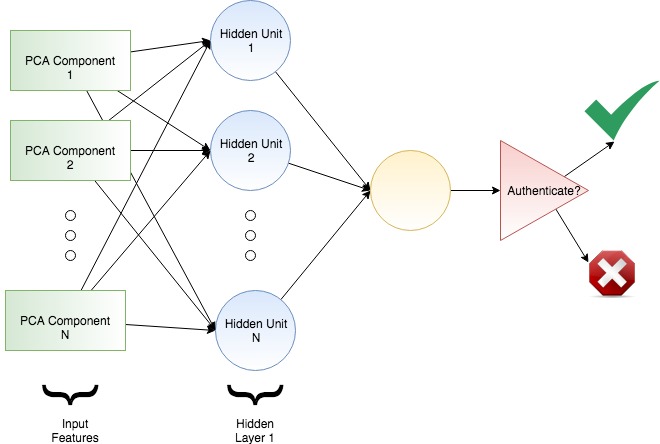}
    \caption{The MLP model used in the proposed system.}
    \label{fig:mlp}
\end{figure}

We used k-Nearest Neighbour (kNN), Random Forest (RF), Multi-Layer Perceptron (MLP) and Linear Discriminant Analysis (LDA)  classifiers which are available in the Weka Machine Learning software \cite{eibe2016weka}. We fine tuned the parameters for different classifiers. The best performing feature set are as follows: N selected as 3 for the kNN, the number of trees is selected as 100 for the random forest and the hidden layer is selected as 1  and hidden unit as 5 for the MLP as shown in Figure \ref{fig:mlp}.    We created a binary authentication model for each user. The selected user's label is set to 1 and others as 0.

We applied 10 fold stratified cross-validation (the distribution of class labels are equal in each fold) for evaluating our system and fine tuned the parameters where 90\% of the dataset is used for training and the rest is used for testing by changing the folds. 

\begin{figure*}[htb!]
    \centering

    \includegraphics[width=0.8\textwidth]{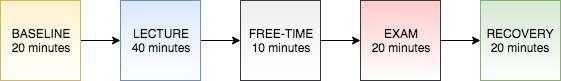}
    \caption{Data collection procedure for the physiological continuous authentication system.}
    \label{fig:datacollection}
\end{figure*} 

\subsection{Evaluation Metrics}

In order to present the results of our authentication system, we provide the performance metrics used in the literature \cite{Vhaduri2019,Sarkar2016, Elkader2018, Zeng2017, Ramli2016} . In authentication systems, there are two types of error which are False Acceptance Rate (FAR) and False Rejection Rate (FRR). These errors are depend on selection of the threshold which can be between 0 and 1 for the ML classifiers. A smaller value will cause a low FAR but high FRR. The point of equilibrium is important for such a system. This point is called Equal Error Rate (EER). The definitions are provided below:

\begin{itemize}
    \item False Acceptance Rate (FAR): It is the ratio of false acceptance divided by the total attempts.
    \item False Rejection Rate (FRR): It is the ratio of denied legitimate attempts to the total number of  attempts.
    \item Equal Error Rate (EER): The common value when FRR and FAR are equal, is called EER \cite{ietbio}. 
\end{itemize}

\section{Data Collection}
\label{experiment}

In this section, we describe the data collection  in real life and the ethics procedure. We collected physiological data from 28 people in controlled real-life settings, during a summer school for teachers. All of the participants are healthy teachers who have no prior medical condition. Before the data collection, subjects received and filled a consent. The gender of participants are 16 male and 12 female, the ages are between 25 and 35. The data collection procedure is shown in Figure \ref{fig:datacollection}. The duration of the total data collection is 110 minutes. The dataset has a baseline (20 minutes), lecture (40 minutes), free-time (10 minutes), examination (20 minutes) and recovery session (20 minutes). We did not use the free-time session which might create a bias on the results. The reason that we had these different scenarios is to create a daily life sequence. A system should take different states into consideration, because HRV can be affected by valance and arousal. During the free-time participants were allowed to take a break from the lecture.   We applied our implementation of Trier Social Stress Test \cite{kirschbaum1993trier} (TSST) which is frequently used for inducing stress.  We selected questions from the mathematics Olympics (which is very hard for the normal population). We told the subjects that this is a test for measuring their intelligence, and we said that a moderate person achieves at least a 75\% score.   Subjects participated in every session and they did not know the objective of the study.  The physiological data is gathered with different brands of commercial smartwatches (8 Empatica E4, 3 Samsung Gear S and 17 Samsung Gear S2). \\

\textit{Ethics}

The procedure of the methodology used in this study is approved by the Institutional Review Board for Research with Human Subjects of Bogaziçi University with the approval number 2018/16. Prior to the data acquisition, each participant received a consent form which explains the experimental procedure and its benefits and implications to both the society and the subject. The procedure was also explained vocally to the subject. The data collection procedure and all of the interventions in this research fully meet the 1964 Declaration of Helsinki \cite{helsinki}. The data is stored anonymously.

\section{Experimental Results and Discussion}
\label{results}
We examined the results in two different subsections. In the first one, we presented and evaluated the authentication results of different devices and the whole system performance. In the second part, by applying a signal data quality filter, we improved the performance of the system. 

\subsection{Effect of Device Type on the Biometric Authentication Performance}
 EER results for all 28 subjects are given  in Table \ref{tab:participants}. These results are calculated by one vs. all tests for all subjects. Average EER results for four different classifiers are presented in Table \ref{tab:average}. We also added the device type and average data quality columns to this table. Data quality presents the non-interpolated percentage of the data after the removal of artifacts. As an example, if the average data quality is 70\%, the remaining 30\% of the data is interpolated. Data quality along with the device type affects the EER results significantly (see Figure \ref{fig:eer}).  We achieved the best performance with Gear S as 98.48\% and 3.96\% EER. The selection of classifier has also an important effect on the EER results. For example, Empatica E4 achieves 19.43\% EER with kNN and 6.77\% with RF classifier. The best classifier is selected as RF in terms of EER. Design of the watch strap as shown in Figure \ref{fig:straps}, PPG sensor quality, built-in processing algorithms of devices might be the factors for the difference in EER results. 

\subsection{Effect of Data Quality Constraint Filter}

\begin{table}
\caption{The EER metrics of each participant with different classifiers.}
\label{tab:participants}

\centering
\begin{adjustbox}{width=0.5\textwidth,center}
\begin{tabular}{@{} l c c c c c c @{}}
\toprule
 Participant/Classifier &kNN& RF&     MLP     &LDA      &Device & Average Quality \\ \midrule
P1  & 20.00 & 1.67 & 3.30 & 5.00 & E4 & 96.00 \\ \midrule
P2 & 22.84 & 5.00 & 10.00 & 8.73 & E4 & 94.00 \\ \midrule
P3 & 5.68 & 3.23 & 5.65 & 9.95 & E4 & 96.00 \\ \midrule
P4 & 31.30 & 12.52 & 15.52 & 15.52 & E4 & 93.00 \\ \midrule
P5 & 30.17 & 9.45 & 27.27 & 27.27 & E4 & 91.50 \\ \midrule
P6 & 23.96 & 8.30 & 15.87 & 19.05 & E4 & 97.00 \\ \midrule
P7 & 17.96 & 11.03 & 22.22 & 25.04 & E4 & 92.00 \\ \midrule
P8 & 3.58 & 2.99 & 4.48 & 5.97 & E4 & 93.00 \\ \midrule
P9 & 4.00 & 5.33 & 5.33 & 6.67 & Gear S & 87.00 \\ \midrule
P10 & 1.33 & 2.67 & 4.00 & 5.33 & Gear S & 95.00 \\ \midrule
P11 & 9.20 & 3.39 & 5.08 & 9.32 & Gear S & 80.00 \\ \midrule
P12 & 37.47 & 18.73 & 23.26 & 21.13 & Gear S2 & 85.00 \\ \midrule
P13 & 30.15 & 10.41 & 21.13 & 22.54 & Gear S2 & 80.00 \\ \midrule
P14 & 13.52 & 3.80 & 8.45 & 9.86 & Gear S2 & 62.00 \\ \midrule
P15 & 31.35 & 4.69 & 12.50 & 17.19 & Gear S2 & 68.00 \\ \midrule
P16 & 18.43 & 6.78 & 20.59 & 20.59 & Gear S2 & 54.00 \\ \midrule
P17 & 36.25 & 18.23 & 27.78 & 29.17 & Gear S2 & 64.00 \\ \midrule
P18 & 43.34 & 24.55 & 43.75 & 40.63 & Gear S2 & 79.00 \\ \midrule
P19 & 21.88 & 2.82 & 8.45 & 11.27 & Gear S2 & 69.00 \\ \midrule
P20 & 34.50 & 11.43 & 18.75 & 20.31 & Gear S2 & 66.00 \\ \midrule
P21 & 39.30 & 19.55 & 21.13 & 22.54 & Gear S2 & 66.00 \\ \midrule
P22 & 25.60 & 13.41 & 18.75 & 18.75 & Gear S2 & 68.00 \\ \midrule
P23 & 31.67 & 14.08 & 23.94 & 21.13 & Gear S2 & 62.00 \\ \midrule
P24 & 33.10 & 14.10 & 23.44 & 31.25 & Gear S2 & 53.00 \\ \midrule
P25 & 20.81 & 7.04 & 15.49 & 19.72 & Gear S2 & 65.00 \\ \midrule
P26 & 33.22 & 16.45 & 24.23 & 23.44 & Gear S2 & 69.00 \\ \midrule
P27 & 37.48 & 20.54 & 28.17 & 29.58 & Gear S2 & 76.00 \\ \midrule
P28 & 36.63 & 19.18 & 23.44 & 26.56 & Gear S2 & 64.00 \\ \bottomrule

\end{tabular}
\end{adjustbox}
\end{table}

\begin{table}
\caption{Authentication performance results, EER values of Empatica E4, Samsung Gear S and Samsung Gear S2 are presented.}
\label{tab:average}
\centering
\begin{adjustbox}{width=0.5\textwidth,center}
\begin{tabular}{@{}lccccc@{}}
\toprule
        Device   & kNN  & RF   & MLP  & LDA & Average Quality   \\ \midrule
Empatica E4 & 19.43\% & 6.77\%  & 13.03\% & 14.56\% & 94.06\% \\
Gear S      & 4.84\%  & 3.96\%  & 4.80\%  & 7.10\% & 87.33\%   \\
Gear S2     & 30.86\% & 13.28\% & 21.36\% & 22.68\% &  67.64\%\\ 
\midrule \midrule
All Devices     & 24.37 \% & 10.08\% & 16.98\% & 18.69 \% &  77.30\% \\ \bottomrule
\end{tabular}
\end{adjustbox}
\end{table}

\begin{figure}
    \centering
    \includegraphics[width=\columnwidth]{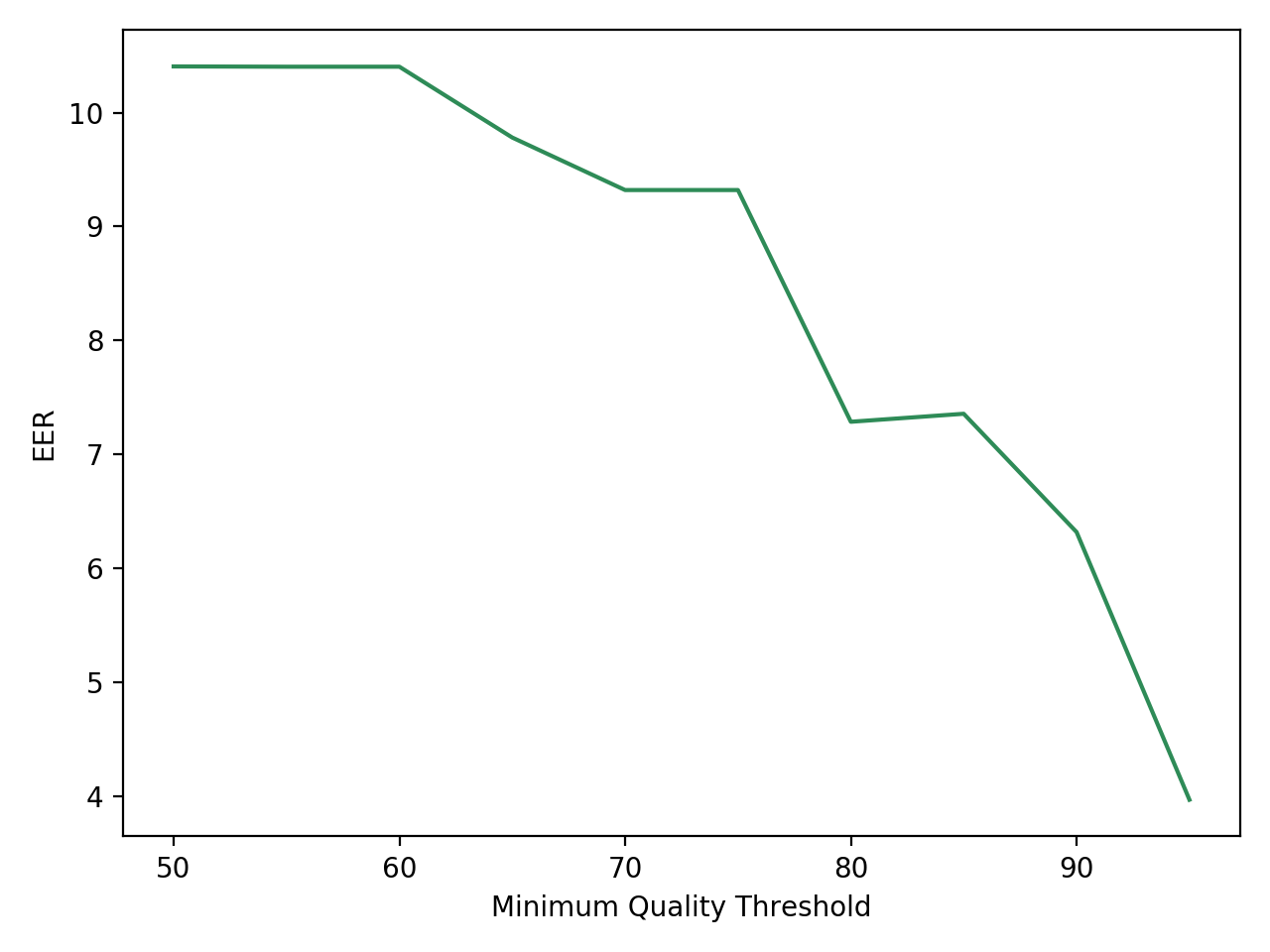}
    \caption{The change of EER metric with respect to minimum quality thresholds. Best achieved EER is 3.96\%. }
    \label{fig:eer}
\end{figure}

In daily life, seamless wrist-worn devices can get noisy signals, which drops the quality of the derived features. It is not possible to collect high-quality data all the time during a day because of various reasons such as high activity level and improper use of smartwatches. After observing that the data quality has a major effect on the authentication performance, we applied a data quality constraint on our data. Suppose that the data quality of a device is 50\%. This means that the other half of the data is obtained by synthetic cubic interpolation data. Therefore, we expect that when the data is compared with other participants' data, it could not be discriminated, because it lost most of the unique characteristics of the PPG data. In Figure \ref{fig:eer}, we evaluated the effect of a quality threshold on EER.  As we investigate the EER results of different device types, Samsung Gear S gives the smallest error rate 3.796\% when compared to Empatica E4 (6.77\%) and Samsung Gear S2 (13.66\%) in low data qualities. As the quality increases, while the error rates of Samsung Gear S (2.67\%) and Empatica E4 (4.4\%) decrease at 95\% quality threshold, Samsung Gear S2 is unable to show the same progress and eventually reaches  18.557\% equal error rate. None of the windows of Samsung Gear S2 has a higher than 95\% quality.  The performance evaluation shows that the proposed system can effectively authenticate with small and consistent error rate which makes it reliable.


\section{Conclusion}
\label{conclusion}
We proposed a scalable, unobtrusive and seamless continuous authentication system with commercial grade smartwatches and smartbands. We collected physiological data from 28 participants and demonstrated the EER measures for each of the participants in a real-life scenario.  We proposed state-of-the art preprocessing for signals coming from real-life data with artifacts due to the physical construction of the smartwatches.    We achieved promising results by using our system (4.4\% EER with Empatica E4).  We showed the effect of different smartwatches. The selection of the classifier for the proposed system is very important. We applied feature based signal processing along with machine learning classifiers (kNN, RF, MLP and LDA). Even-though, Gear S2 is a newer model of Gear S, due to its leather strap, the signals coming from the heart rate monitoring unit contained higher amount of artifacts, therefore it affects the overall quality of the RR intervals and the authentication system's performance. For the authentication systems based on PPG sensors, sport straps can be a better choice, as shown in Figure \ref{fig:straps}.    We showed that HRV can be used for continuous authentication without interrupting the activity of the user. We applied a signal removal procedure by using the overall RR interval quality measure, a higher quality leads to better performance after 80\% quality threshold.  The performance of the scheme varies between individuals. This conclusion is aligned with the literature \cite{Vhaduri2019, Ramli2016}.  The minimum required amount of recording to apply our system for authentication is 2 minutes, once that is satisfied, authentication can be validated in seconds thanks to the sliding window approach. It logouts the user, once he/she leaves off the watch. Our system can be implemented on any wrist-worn device which can provide RR intervals without a need for the raw PPG. The proposed methodology can be used with various applications requiring continuous authentication.  This study also has some limitations. The performance of the system on the data coming from different days is still unknown. As future works, we plan to apply our system completely in the wild settings with more participants and longer physiological recordings and show the performance of the framework. All of the evaluations are done in the same context, therefore in different types of contexts, the system might achieve better performance.

\ifCLASSOPTIONcaptionsoff
  \newpage
\fi



\bibliography{bare_jrnl_compsoc}{}
\bibliographystyle{ieeetr}

%



%


\begin{IEEEbiography}[{\includegraphics[width=1in,height=1.25in,clip,keepaspectratio]{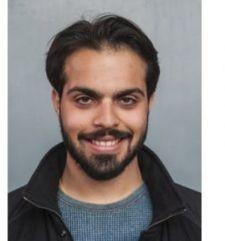}}]{Deniz Ekiz} received the MS degree from Computer Engineering Department, Bogazici University, Turkey in 2019. He 
 is a Ph.D. candidate in the Computer
Engineering Department of Bogaziçi University, Turkey.
His research is focused on the health-related
applications of wearable technology
\end{IEEEbiography}


\begin{IEEEbiography}[{\includegraphics[width=1in,height=1.25in,clip,keepaspectratio]{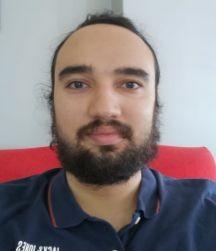}}]{Yekta Said Can}
studied Computer Engineering in
 Bogazici University,
Istanbul, Turkey where he obtained his B.Sc Degree
in 2012. He obtained his M.Sc degree in the same
department, in 2014 while working as a researcher
at TUBITAK BILGEM for two years. He is
pursuing a PhD degree right now in Computer
Engineering at Bogazici University. His research interest includes
watermarking, speech and speaker recognition, physiological signal
processing and machine learning.
\end{IEEEbiography}

\begin{IEEEbiography}[{\includegraphics[width=1in,height=1.25in,clip,keepaspectratio]{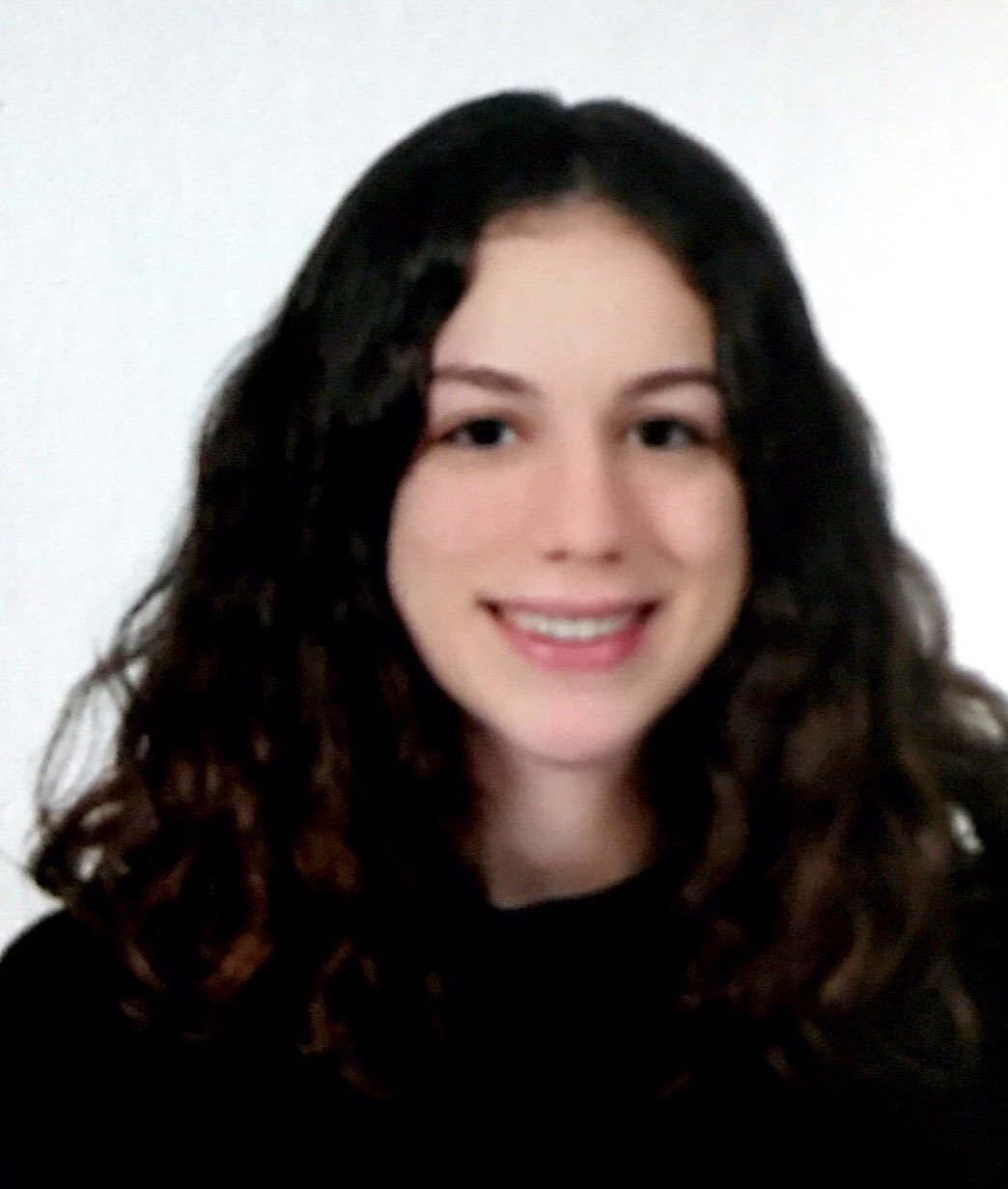}}]{Yagmur Ceren Dardagan}
is a senior student in Computer Engineering Department of Bogazici University, Turkey. Her research interests include physiological signal processing, machine learning and pervasive health applications.
\end{IEEEbiography}

\begin{IEEEbiography}[{\includegraphics[width=1in,height=1.25in,clip,keepaspectratio]{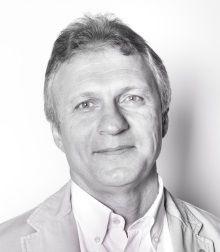}}]{Cem Ersoy}
received the Ph.D. degree from
Polytechnic University, New York in
1992. He was an R\&D Engineer in NETAS A. S.
from 1984 to 1986. He is a Professor of computer
engineering at Bogazici University, Turkey. He is also the
Vice Director of the Telecommunications and
Informatics Technologies Research Center,
TETAM. His research interests include
wireless/cellular/ad-hoc/sensor networks, activity
recognition, and ambient intelligence for pervasive
health applications, green 5G and beyond networks, and mobile cloud/edge/fog
computing. He was the Chairman of the IEEE
Communications Society Turkish Chapter  eight years. He is a
member of the IFIP.
\end{IEEEbiography}



\end{document}